\documentclass[twocolumn,prl,showpacs,superscriptaddress,amssymb,amsmath,amsmath]{revtex4-1}
\usepackage{graphicx}
\usepackage{epstopdf}
\usepackage{bm}
\usepackage{hyperref}
\usepackage{wasysym}
\usepackage{comment}

\hypersetup{%
   pdfpagemode=None, 
   pdfstartpage=1,
   pdfmenubar=true,
   pdftoolbar=true,
   colorlinks = true,
   linkcolor=blue,
   citecolor=blue,
   urlcolor=blue,
   bookmarksopen=false
 }

\newcommand{\be}{\begin{equation}}
\newcommand{\ee}{\end{equation}}

\begin{document}
\title{\textit{Ab initio} properties of the NaLi molecule in the $a^3\Sigma^+$ electronic state}

\author{Marcin Gronowski}
\affiliation{Faculty of Physics, University of Warsaw, Pasteura 5, 02-093 Warsaw, Poland}
\author{Adam M. Koza}
\affiliation{Faculty of Physics, University of Warsaw, Pasteura 5, 02-093 Warsaw, Poland}
\affiliation{Faculty of Chemistry, University of Warsaw, Pasteura 1, 02-093 Warsaw, Poland}
\author{Micha\l~Tomza}
\email{michal.tomza@fuw.edu.pl}
\affiliation{Faculty of Physics, University of Warsaw, Pasteura 5, 02-093 Warsaw, Poland}

\date{\today}

\begin{abstract}
Ultracold polar and magnetic ${}^{23}$Na${}^6$Li molecules in the rovibrational ground state of the lowest triplet $a^3\Sigma^+$ electronic state have been recently produced. Here, we calculate the electronic and rovibrational structure of these 14-electron molecules with spectroscopic accuracy ($<0.5\,$cm$^{-1}$) using state-of-the-art \textit{ab initio} methods of quantum chemistry. We employ the hierarchy of the coupled-cluster wave functions and Gaussian basis sets extrapolated to the complete basis set limit. We show that the inclusion of higher-level excitations, core-electron correlation, relativistic, QED, and adiabatic corrections is necessary to reproduce accurately scattering and spectroscopic properties of alkali-metal systems. We obtain the well depth, $D_e=229.9(5)\,$cm$^{-1}$, the dissociation energy, $D_0=208.2(5)\,$cm$^{-1}$, and the scattering length, $a_s=-84^{+25}_{-41}\,$bohr, in good agreement with recent experimental measurements. We predict the permanent electric dipole moment in the rovibrational ground state, $d_0=$0.167(1)$\,$debye. These values are obtained without any adjustment to experimental data, showing that quantum chemistry methods are capable of predicting scattering properties of many-electron systems, provided relatively weak interaction and small reduced mass of the system.

\end{abstract}

\maketitle

\paragraph{Introduction.} The realization of ultracold gases of atoms and molecules has allowed for numerous unprecedented experiments probing quantum phenomena in physics and chemistry~\cite{GrossScience17,BohnScience17}. Ultracold molecules such as KRb~\cite{NiScience08}, RbCs~\cite{TakekoshiPRL14,MolonyPRL14}, NaK~\cite{ParkPRL15}, NaRb~\cite{GuoPRL16} have been produced in their ground rovibrational and electronic states and employed in ground-breaking experiments on controlled chemical reactions~\cite{OspelkausScience10,NiNature10,MirandaNatPhys11,YangScinece19,GregoryNC19} and quantum many-body dynamics~\cite{BoNature13}.
Recently, NaLi molecules in the ground rovibrational level of the lowest triplet electronic state have been created and investigated ~\cite{HeoPRA12,RvachovPRL17,RvachovPCCP18a,RvachovPCCP18b,Son2019}. These molecules possess both electric and magnetic dipole moments~\cite{TomzaPRA13}, making them promising candidates for various applications in the quantum simulation of many-body physics~\cite{MicheliNatPhys06}. On the other hand, high precision spectroscopy of ultracold molecules allows for tests of fundamental theories~\cite{SalumbidesPRD13,ACMENature18}, while ultracold molecular collisions can be useful to probe intermolecular interactions~\cite{LavertNatChem14,TomzaPRL15}.

The electronic structure calculations for few-electron atoms and molecules (up to four electrons) have reached unparalleled accuracy. The nonrelativistic electronic Schr\"odinger equation can be solved almost exactly for such systems, while relativistic, quantum electrodynamics (QED), adiabatic, and nonadiabatic corrections can be included in a systematic and controlled way~\cite{KolosJCP68,PrzybytekJCP05,PiszczatowskiJCTC09,PrzybytekPRL10,PuchalskiPRL16,PrzybytekPRL17,PuchalskiPRL18,PuchalskiPRL19}. Calculations of the interaction potential well depth for the simplest H$_2$~\cite{PuchalskiPRL19} and He$_2$~\cite{PrzybytekPRL17} molecules have achieved uncertainties as small as $2\times 10^{-5}\,$cm$^{-1}$ (1$\,$ppb) and $10^{-4}\,$cm$^{-1}$ (20$\,$ppm), respectively. High accuracy has also been presented for systems involving weak interactions with light H$_2$ or He, e.g.~H$_2$+CO~\cite{JankowskiScience12}, H$_2$+He~\cite{BakrJCP13,KleinNP17}, He+Rb~\cite{KnoopPRA14}, and H$_2$+Li~\cite{MakridesPRA19}. Recently, the six-electron Li$_2$~\cite{LesiukPhD} and eight-electron Be$_2$~\cite{LesiukJCTC19} molecules were investigated, including leading relativistic and QED corrections, with the uncertainty of the well depth of 0.3$\,$cm$^{-1}$ (0.9$\permil$) and 2.5$\,$cm$^{-1}$ (3$\permil$), respectively.

Despite many successes in the theoretical description of few-electron molecules as well as ultracold atomic and molecular gases, the interatomic interactions between ultracold alkali-metal or alkaline-earth-metal atoms have never been described accurately enough to predict the scattering length solely based on \textit{ab initio} electronic structure calculations without any adjustment to experimental data~\cite{ChinRMP10}. 
The main reason is the high sensitivity of the scattering length to the accuracy of the interaction potential and especially to the position of the last weakly bound rovibrational state~\cite{ChinRMP10}. In turn, the computational costs of calculating exactly interaction potentials increase factorially with the number of involved electrons due to the related factorial growth of many-electron wave functions and Hilbert spaces~\cite{Helgaker}.

Here, we report accurate calculations of the electronic and rovibrational structure of the ${}^{23}$Na${}^6$Li molecule in the triplet $a^3\Sigma^+$ electronic state. NaLi contains 14 electrons. Despite the large number of electrons, we reach spectroscopic accuracy ($<0.5$cm$^{-1}$) using state-of-the-art \textit{ab  initio} methods of quantum chemistry, including the coupled-cluster wave functions and Gaussian single-particle basis sets extrapolated to the complete basis set limit. We predict the dissociation energy of 208.2(5)$\,$cm$^{-1}$ for the ground triplet-state ${}^{23}$Na${}^6$Li molecule and the scattering length of $-84^{+25}_{-41}\,$bohr for the spin-polarized ${}^{23}$Na+${}^6$Li collisions, both in good agreement with experimental data~\cite{SchusterPRA12,SteinkePRA12,RvachovPRL17,RvachovPCCP18b}. We also obtain the permanent electric dipole moment of 0.167(1)$\,$debye for the rovibrational ground state. We calculate and show that the inclusion of higher-level excitations, core-electron correlation, relativistic, QED, and adiabatic corrections is necessary to reproduce accurately considered scattering and spectroscopic properties. We demonstrate that quantum chemistry methods are capable of predicting scattering properties of many-electron systems without any adjustment to experimental data. Finally, we show that the recent experimental potential obtained as a fit to a highly accurate vibrational spectrum~\cite{RvachovPCCP18b} has a less accurate shape due to using an inaccurate theoretical equilibrium distance and thus may not reproduce the rotational spectrum correctly. 

\begin{figure}[tb]
\begin{center}
\includegraphics[width=\columnwidth]{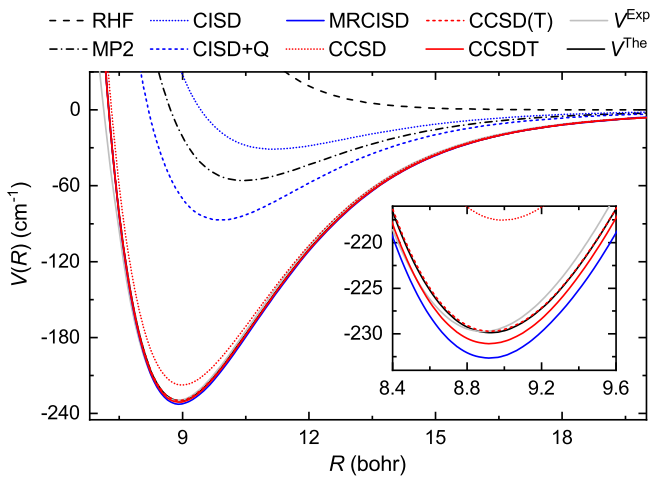}
\end{center}
\caption{Interaction energy as a function of the internuclear distance for the $a^3\Sigma^+$ electronic state of the NaLi molecule from nonrelativistic Born-Oppenheimer calculations at the RHF, MP2, CISD, CISD+Q, MRCISD, CCSD, CCSD(T), and CCSDT levels of theory. The inset shows an enlarged part around the equilibrium distance. The experimental $V^\text{Exp}$~\cite{RvachovPCCP18b} and total present $V^\text{The}$ potentials are plotted for comparison. See the text for details.} 
\label{fig:curves}
\end{figure}

\paragraph{Electronic structure.} The exact many-electron calculations using the full configuration interaction (FCI) method are infeasible for 14-electron molecules~\cite{Helgaker}. Therefore to obtain the potential energy curve (PEC) for the ${}^{23}$Na${}^6$Li molecule in the $a^3\Sigma^+$ electronic state within the Born-Oppenheimer approximation, we employ the composite approach and calculate different contributions to the electronic interaction energy, $V_\text{int}$, separately~\cite{RezacCR16}
\begin{equation}
V_\text{int}=V_\text{BO}^\text{CCSD(T)}+\delta V_\text{BO}^\text{TQP}+\delta V_\text{rel}+\delta V_\text{QED}+\delta V_\text{ad}\,,
\end{equation}
where each term is computed using the highest level of available theory. $V_\text{BO}^\text{CCSD(T)}$ is the leading part of the interaction energy obtained using the restricted open-shell coupled cluster method including single, double, and noniterative triple excitations, RCCSD(T)~\cite{KnowlesJCP93}. $\delta V_\text{BO}^\text{TQP}$ stands for the contribution from the inclusion of higher-level excitations in the coupled-cluster wave function~\cite{KallayJCP01}. $\delta V_\text{rel}$ counts for the leading relativistic effects ($\sim\alpha^2$)~\cite{KlopperJCC97}. $\delta V_\text{QED}$ estimates the leading quantum electrodynamics (QED) correction ($\sim\alpha^3$)~\cite{PyykkoPRA01}. $\delta V_\text{ad}$ is the diagonal adiabatic (Born-Oppenheimer) correction~\cite{HandyJCP86}. The $a^3\Sigma^+$ electronic state is the first excited but the lowest-energy triplet state of the NaLi molecule, therefore in all calculations we can employ methods implemented for the ground state of a given spin multiplicity.

The orbitals of Na and Li atoms are constructed using the augmented correlation-consistent polarized core-valence Gaussian basis sets, aug-cc-pCVnZ with $n={T,Q,5}$~\cite{PrascherTCA10}, additionally augmented by the set of the $[3s3p2d1f1g]$ bond functions (BF)~\cite{midbond}. We also develop and employ larger basis sets of quality approaching aug-cc-pCV6Z and optimized for the ground-state interactions in the Li$_2$ and Na$_2$ dimers. The interaction energies are obtained with the supermolecular method with the basis set superposition error (BSSE) corrected by using the counterpoise correction~\cite{BoysMP70}. The energies and interaction-induced properties are appropriately extrapolated to the complete basis set (CBS) limit, using $n^{-3}$ scheme~\cite{HelgakerJCP97}. The nonrelativistic calculations are performed with the Molpro and MRCC packages of \textit{ab initio} programs~\cite{molpro2,molpro,KallayJCP20}, while the relativistic and adiabatic corrections are calculated using the CFOUR and DIRAC packages~\cite{cfour,DIRAC18}. The uncertainties of calculated energies are estimated at each internuclear distance based on the analysis of the convergence with the size of employed basis sets and wave functions. The uncertainties due to the basis-set incompleteness are estimated as the difference between the values extrapolated to the CBS limit~\cite{HelgakerJCP97} and the largest employed basis sets or as the difference between the values for the two largest basis sets. The uncertainties due to wave-function and Hamiltonian incompleteness are conservatively estimated based on known values for smaller systems or lower-order terms~\cite{PuchalskiPRL19,PrzybytekPRL17,LesiukJCTC19}. The leading $V_\text{BO}^\text{CCSD(T)}$ term is calculated for 120 distances between 6$\,$bohr and 50$\,$bohr (every 0.1$\,$bohr around the equilibrium distance), while other terms are calculated for 50 distances~\footnote{See Supplemental Material at http://link.aps.org/supplemental/XXXX for the calculated interaction energies in a numerical form.}. Next, they are interpolated using the cubic spline method, which is used instead of fitting analytical functions to avoid fitting errors.  

\paragraph{Nonrelativistic interaction potential.} The dominating part of the nonrelativistic interaction energy can be decomposed into the Hartree-Fock (mean-field) and correlation parts~\cite{Helgaker}. The restricted Hartree-Fock (RHF) contribution to the well depth at the equilibrium distance of the final interaction potential, $R_e=8.924$, is $207.46(1)\,$cm$^{-1}$. The positive sign means that the NaLi system in the $a^3\Sigma^+$ electronic state is not bound when beyond mean-field corrections are not included (see Fig.~\ref{fig:curves}). To calculate the correlation energy we use the hierarchy of the coupled cluster methods~\cite{BartlettRMP07} and its final contribution to the well depth at $R_e$ is $V_e^\text{corr}=-438.7(3)\,$cm$^{-1}$.

To analyze the convergence of the interaction energy calculation, we present results obtained with several methods in Fig.~\ref{fig:curves}. The second-order many-body (M{\o}ller-Plesset) perturbation theory (MP2)~\cite{BartlettARPC81} reproduces 51.0$\,$\% of $V_e^\text{corr}$. The configuration interaction method including single and double excitations (CISD)~\cite{Helgaker} reproduces only 39.6$\,$\% of $V_e^\text{corr}$, while inclusion of the Davidson correction (CISD+Q)~\cite{Helgaker} improves it to 62.5$\,$\%. The coupled cluster method including single and double excitations (CCSD)~\cite{BartlettRMP07} reproduces 96.8$\,$\% of $V_e^\text{corr}$, while the inclusion of noniterative triple excitation (CCSD(T))~\cite{BartlettRMP07} improves it to 99.6$\,$\%. The multireference configuration interaction method including single and double excitations (MRCISD)~\cite{WernerJCP88} reproduces 100.4$\,$\% of $V_e^\text{corr}$. The poor performance of the CISD method, as compared to CCSD and MRCISD, indicates a significant contribution from the interaction between electron-correlated parts of the atomic wave functions.

\begin{figure}[tb]
\begin{center}
\includegraphics[width=\columnwidth]{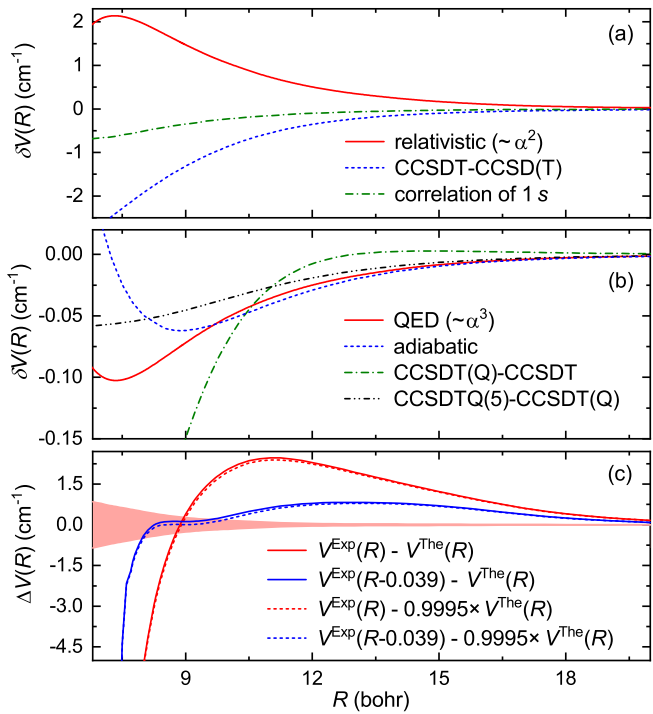}
\end{center}
\caption{(a),(b) Higher-level-excitation, core-electron-correlation, relativistic, QED, and adiabatic corrections to the Born-Oppenheimer interaction energy. (c) Difference between the experimental $V^\text{Exp}(R)$~\cite{RvachovPCCP18b} and present $V^\text{The}(R)$ PECs. Shaded area represents the uncertainty of the present PEC. See the text for details.}
\label{fig:corr}
\end{figure}

The final well depth of the nonrelativistic interaction potential within the Born-Oppenheimer approximation is $-231.3(3)\,$cm$^{-1}$. The correlation of $1s$ electrons of Na accounts for $-0.36\,$cm$^{-1}$ at $R_e$. The noniterative triple excitations in the CCSD(T) method contributes $-12.35\,$cm$^{-1}$, while the remaining part of the triple excitations accounts for $-1.46(5)\,$cm$^{-1}$. The noniterative quadruple excitations in the CCSDST(Q) method contributes $-0.16(5)\,$cm$^{-1}$. The estimated contribution of quintuple and higher-level excitations from the CCSDSTQ(P) method is $-0.05(5)\,$cm$^{-1}$. Calculations with the RHF, MP2, CISD, MRCISD, CCSD, and CCSD(T) methods employ the aug-cc-pCV6Z basis set augmented by the midbond functions to accelerate convergence toward the CBS limit. Calculations with the CCSDT method use the aug-cc-pCV$n$Z basis sets extrapolated to the CBS limit. Calculations with the CCSDT(Q) and  CCSDTQ(P) methods use the aug-cc-pCVTZ and aug-cc-pVDZ basis sets, respectively.

\paragraph{Relativistic correction.} The leading relativistic effects, proportional to $\alpha^2$~\cite{Dyall2007}, are included using the second-order direct perturbation theory (DPT)~\cite{KlopperJCC97}, which is equivalent to the inclusion of the mass-velocity, one-electron Darwin, and two-electron Darwin terms in the Breit-Pauli approximation~\cite{CowanJOSA76}. The remaining Breit (spin-spin and orbit-orbit) term is approximated by the Gaunt correction~\cite{GauntPRCL29,GYORGYMP01}, while the spin-orbit term is zero for the $^3\Sigma$ state. The DPT calculation employs the CCSD method and aug-cc-pCV$n$Z basis sets extrapolated to the CBS limit. The Gaunt correction is calculated as a difference between the interaction energies obtained with the four-component Dirac-Coulomb-Gaunt and Dirac-Coulomb Hamiltonians~\cite{TrondCPC11} with the wave function at the RHF level~\cite{VisscherADNT97} and all-electron relativistic quadruple-zeta basis sets~\cite{DyallTCA16}.
The calculated total relativistic correction is plotted in Fig.~\ref{fig:corr}(a) and takes the value of 1.51(10)$\,$cm$^{-1}$ at $R_e$, which is larger than the uncertainty of the the calculated nonrelativistic potential. The reported uncertainty counts for both the numerical uncertainty of the present calculation and the lack of inclusion of higher-order terms in the Hamiltonian and wave function. The mass-velocity term contributes 4.71$\,$cm$^{-1}$, one- and two-electron Darwin terms contributes -3.39$\,$cm$^{-1}$ and 0.05$\,$cm$^{-1}$, while the Gaunt correction contributes 0.14$\,$cm$^{-1}$ of $\delta V_\text{rel}$ at~$R_e$. 

\paragraph{QED correction.} The leading quantum electrodynamics effects, i.e.~the Lamb shift proportional to $\alpha^3$ and $\alpha^3\ln\alpha$~\cite{PyykkoPRA01}, are estimated using the molecular one-electron Darwin term $D_1$ obtained in the Breit-Pauli approximation, as described in the previous paragraph, and Bethe logarithm $\ln k_0$~\cite{PyykkoPRA01,LesiukJCTC19}. The molecular Bethe logarithm is approximated by $\ln k_0^\text{NaLi}=(\ln k_0^\text{Li}D_1^\text{Li}+\ln k_0^\text{Na}D_1^\text{Na})/(D_1^\text{Li}+D_1^\text{Na})$~\cite{PiszczatowskiJCTC09}, where the atomic Bethe logarithms are $\ln k_0^\text{Li}=5.17817(3)$ and $\ln k_0^\text{Na}=7.7845$~\cite{PachuckiPRA03,Lesiuk}. Thus, the dominant one-electron contribution to the molecular Lamb shift is given by $\delta V_\text{QED}=\frac{8\alpha}{3\pi}(\frac{19}{30}-2\ln\alpha-\ln k_0^\text{NaLi})D_1^\text{NaLi}(R)$. The calculated QED correction is plotted in Fig.~\ref{fig:corr}(b) and takes the value of -0.05(4)$\,$cm$^{-1}$ at $R_e$. The reported uncertainty counts for both the numerical uncertainty of the present calculation and the lack of inclusion of higher-order terms.

\paragraph{Adiabatic correction.}
The diagonal adiabatic (Born-Oppenheimer) correction, which is the leading correction beyond the Born-Oppenheimer approximation, is calculated using the first-order perpetration theory~\cite{HandyJCP86} with the wave function at the CCSD level~\cite{GaussJCP06} and aug-cc-pCV5Z basis sets. The adiabatic correction is presented in Fig.~\ref{fig:corr}(b) and takes the value of 0.06(2)$\,$cm$^{-1}$ at $R_e$. Additionally, we conservatively estimate the uncertainty due to the unknown nonadiabatic corrections to have a value of $30\,\%$ of the calculated adiabatic correction~\cite{PachuckiJCP15}.

\paragraph{Total interaction potential.} Finally, the well depth of PEC for the $a^3\Sigma^+$ electronic state is $D_e=229.9(5)\,$cm$^{-1}$ and the equilibrium distance is $R_e=8.924(6)\,$bohr. Present results are more than an order of magnitude more accurate than previous theoretical results for the potential well depth~\cite{SchmidtCPL84,MabroukJPB08,PetsalakisJCP08,MieszczaninMP14}. The present value of the well depth agrees well with analytical fit to experimental data $D_e$=229.753$\,$cm$^{-1}$~\cite{RvachovPCCP18b}, however the experimental fit used the rovibrational levels with $j=0$ and the older less-accurate theoretical value of $R_e$=8.88463$\,$bohr. Thus, despite the experimental PEC~\cite{RvachovPCCP18b} reproduces the potential volume and measured vibrational energies accurately, its shape and associated rotational energies may be less accurate. In Fig.~\ref{fig:corr}(c), we plot the difference between the experimental and present PECs. The difference is larger than present uncertainties except for distances around $R_e$. Using the present $R_e$ in the experimental fit (i.e.,~the shift of 0.039$\,$bohr) reduces the difference significantly, which still is larger than the present uncertainties. The experimental potential is too deep for $R<R_e$ and too shallow for $R>R_e$.  

\begin{table}[tb]
\caption{Well depth $D_e$, dissociation energy $D_0$, and scattering length $a_s$ for ${}^{23}$Na+${}^6$Li in the $a^3\Sigma^+$ electronic state obtained with potentials at different levels of theory.\label{tab:spec}} 
\begin{ruledtabular}
\begin{tabular}{lrrr}
Potential & $D_e\,$(cm$^{-1}$) & $D_0\,$(cm$^{-1}$) & $a_s\,$(bohr) \\
\hline
$V_\text{BO}^\text{CCSD(T)}$ &  229.60(15) & 207.96(14) & $-78^{+15}_{-19}$ \\
$V_\text{BO}^\text{CCSD(T)}$+$\delta V_\text{BO}^\text{T}$ & 231.06(20) & 209.34(19) & $-143^{+35}_{-57}$ \\
$V_\text{BO}^\text{CCSD(T)}$+$\delta V_\text{BO}^\text{T}$ +$\delta V_\text{BO}^\text{QP}$ & 231.27(30) & 209.53(28) & $-155^{+43}_{-77}$  \\
$V_\text{BO}$+$\delta V_\text{rel}$ & 229.76(40) & 208.10(38) & $-81^{+21}_{-34}$  \\
$V_\text{BO}$+$\delta V_\text{rel}$+$\delta V_\text{QED}$  & 229.81(44) & 208.15(42) & $-82^{+23}_{-37}$   \\
$V_\text{BO}$+$\delta V_\text{rel}$+$\delta V_\text{QED}$+$\delta V_\text{ad}$ & 229.87(48) & 208.21(46) & $-84^{+25}_{-41}$\\
Exp.~\cite{SchusterPRA12} & - & - & $-76\pm$5  \\
Exp.~\cite{RvachovPCCP18b} & 229.753 & 208.0826(3) & $-74$\\
\end{tabular}
\end{ruledtabular}
\end{table}

\paragraph{Scattering length.} The calculated PEC for the $a^3\Sigma^+$ electronic state describes ground-state spin-polarized $^{23}$Na+$^6$Li collisions. In the ultracold regime, scattering is fully characterized by the $s$-wave scattering length $a_s$, which is a crucial parameter for all ultracold physics experiments and which is highly sensitive to the accuracy of the PEC volume and especially to the position of the last weakly bound state~\cite{ChinRMP10}. We calculate this property by solving numerically exactly the Schr\"odinger equation for the nuclear motion within the $S$-matrix formalism~\cite{SchmidPRL18,TomzaPRL14}.
Atomic masses are used~\cite{nist}. The calculated PES is connected at $R=40\,$bohr with the long-range multipole expansion of the dispersion interaction given by $C_6=1467(2)\,$a.u., $C_8=98800(1100)\,$a.u., and $C_{10}=9.16\times 10^6\,$a.u.~coefficients~\cite{DereviankoPRA01,PorsevJCP03}. The long-range expansion agrees within 1\% with the calculated PEC for $R=30$-$40\,$bohr with the difference $<0.001\,$cm$^{-1}$ at $R=40\,$bohr. The numerical uncertainty of scattering length calculation contributes less than 1$\,$bohr of its value. Thus, the uncertainty of the calculated PEC is the main source of the uncertainty of the scattering length. 

The scattering length obtained with PEC calculated at different levels of theory is presented in Table~\ref{tab:spec}. Surprisingly, due to the accidental error cancellation, already the nonrelativistic potential with the CCSD(T) method gives $-78^{+15}_{-19}$ close to the experimental value of $-76(5)$~\cite{SchusterPRA12}. Inclusion of the full triple excitations changes $a_s$ by $-65(4)\,$bohr, while higher-level excitations contribute $-12(6)\,$bohr. The leading relativistic correction changes $a_s$ by $74(5)\,$bohr. Finally, the leading QED and adiabatic corrections contribute $-2(1)\,$bohr and $-2(1)\,$bohr, respectively. The final scattering length is $-84^{+25}_{-41}\,$bohr. 

All considered corrections to the leading CCSD(T) interaction potential change the scattering length by values larger than typical uncertainties of experimental results extracted from ultracold collision measurements. The importance of the relativistic effects on ultracold Na+Li collisions is in sharp contrast to recent results on cold NO+He collisions, where relativistic effects were shown to be negligible~\cite{JonghScience20}. This highlights the significance of proper inclusion of higher-level-excitation, core-electron-correlation, relativistic, QED, and adiabatic corrections. Careful estimation of their theoretical uncertainties is also important to avoid unjustified implications because of an accidental agreement with experimental results due to an accidental error cancellation.

\begin{table}[tb]
\caption{Calculated vibrational binding energies $E_v^\text{The}$ of the ${}^{23}$Na${}^6$Li molecule in the $a^3\Sigma^+$ electronic state (in cm$^{-1}$) compared to the experimental results $E_v^\text{Exp}$ from Ref.~\cite{RvachovPCCP18b}.\label{tab:vib}} 
\begin{ruledtabular}
\begin{tabular}{lrrrrr}
$v$ & $E_v^\text{The}$ & $E_v^\text{Exp}$ & $E_{v}^\text{The}-E_{v+1}^\text{The}$ & $E_{v}^\text{Exp}-E_{v+1}^\text{Exp}$  \\
\hline
0  & 208.2(5)  & 208.0826(3) & 40.19(3) & 40.172(1) \\
1  & 168.0(4)  & 167.910(1)  & 36.01(4) & 35.988(3)  \\
2  & 132.0(4)  & 131.922(2)  & 31.77(4) & 31.753(3) \\
3  & 100.2(3)  & 100.169(1)  & 27.45(4) & 27.440(3) \\
4  & 72.9(3)   & 72.730(2)   & 23.03(4) & 23.008(3) \\
5  & 49.8(3)   & 49.722(1)   & 18.53(4) & 18.512(2) \\
6  & 31.2(2)   & 31.210(1)   & 13.98(5) & 13.960(2) \\
7  & 17.3(2)   & 17.250(1)   & 9.49(6)  & 9.482(2) \\
8  & 7.77(12)  & 7.768(1)    & 5.36(6)  & 5.354(2) \\
9  & 2.42(6)   & 2.4137(3)   & 2.10(5)  & 2.1018(7) \\
10 & 0.314(17) & 0.3119(3)   & - & - \\
\end{tabular}
\end{ruledtabular}
\end{table}

\paragraph{Vibrational structure.} There are 11 vibrational levels ($v,j$=0) supported by the $a^3\Sigma^+$ electronic state of the $^{23}$Na$^6$Li molecule~\cite{RvachovPCCP18b}. We calculate these vibrational levels using numerically exact diagonalization of the Hamiltonian for the nuclear motion within the discrete variable representation (DVR) on the nonequidistant grid~\cite{TiesingaPRA98}. The dissociation energy of the ground rovibrational level ($v$=0,$j$=0) obtained with the most accurate PEC is $D_0$=208.2(5)$\,$cm$^{-1}$ in good agreement with the experimental value of 208.0826(3)$\,$cm$^{-1}$. The convergence of the dissociation energy with the small contributions to the interaction energy is presented in Table~\ref{tab:spec}. 

Calculated binding energies of all 11 vibrational levels, together with their experimental values~\cite{RvachovPCCP18b}, are collected in Table~\ref{tab:vib}. The root-mean-square deviation (RMSD) of calculated values from the experimental ones is 0.07$\,$cm$^{-1}$ (0.2$\,$\%). We also calculate the vibrational excitation energies $E_{v}-E_{v+1}$, which RMSD is 0.013$\,$cm$^{-1}$ (0.07$\,$\%). The calculated fundamental vibrational excitation energy is 40.19(3)$\,$cm$^{-1}$ in good agreement with the experimental value of 40.172(1)$\,$cm$^{-1}$. The vibrationally averaged interatomic distance $\langle R\rangle_v$ for the most deeply and weekly bound vibrational levels is 9.139(3)$\,$bohr and 29.4(3)$\,$bohr, respectively.

The PEC from Ref.~\cite{RvachovPCCP18b}, which is the fit to the experimental data, reproduces vibrational binding and excitation energies with RMSD of 0.024$\,$cm$^{-1}$ and 0.023$\,$cm$^{-1}$. If the present PEC is uniformly rescaled by 0.9995, then it reproduces experimental vibrational binding and excitation energies with RMSD of 0.010$\,$cm$^{-1}$ and 0.006$\,$cm$^{-1}$, respectively. The corresponding scattering length is $-80\,$bohr. Such a rescaling procedure does not reduce the difference between the present and experimental PECs in Fig.~\ref{fig:corr}(c) significantly, highlighting that the present theoretical shape of PEC with the well-estimated uncertainties should be more accurate than the experimental one.

\paragraph{Rotational structure.} We also calculate the rovibrational states for non-zero rotational angular momenta. The two lowest rotational excitation energies are 
\begin{equation*}
\begin{split}
E_{(v=0,j=0)\to(v=0,j=1)}& = 9.231(6)\,\text{GHz}\,,\\
E_{(v=0,j=0)\to(v=0,j=2)}& = 27.688(18)\,\text{GHz}\,.
\end{split}
\end{equation*}
The second value agrees well with the sole experimental value of 27.7(1)$\,$GHz~\cite{RvachovPCCP18b}. The calculated rotational constant is $B_0=4.616(3)\,$GHz. The present theoretical rotational excitation energies are more accurate than the recent experimental measurement~\cite{RvachovPCCP18b} and can guide future spectroscopic studies. We predict the existence of 243 rovibrational levels in total.

\paragraph{Static electric dipole moment and polarizability.} Intermolecular interactions and interactions with an external electric field are governed, in the first order, by the molecular permanent electric dipole moment, which for the rovibrational ground state of the ${}^{23}$Na${}^6$Li molecule in the $a^3\Sigma^+$ state, we calculate to be 0.167(1)$\,$debye in the molecular frame oriented from Li to Na. In the second order of perturbation theory, the isotropic and anisotropic components of the molecular polarizability tensor are important, which for the rovibrational ground state, we predict to be 364(3)$\,$a.u. and 273(3)$\,$a.u (corresponding parallel and perpendicular components are 546(3)$\,$a.u. and 273(3)$\,$a.u). The presented static electric properties are obtained with the finite field approach and the CCSD(T) method with the relativistic Douglas-Kroll-Hess correction~\cite{Dyall2007} and aug-cc-pCV6Z basis sets, and they agree well with previous theoretical results~\cite{AymarJCP05,DeiglmayrJCP08,TomzaPRA13}.

\paragraph{Other small corrections.} Because the studied electronic state has a non-zero electronic spin, three other small corrections contribute to the rovibrational spectrum of the ${}^{23}$Na${}^6$Li molecule in the $a^3\Sigma^+$ state. The interaction-induced variation of the hyperfine coupling between electronic and nuclear spins is expected to be smaller than 0.0005$\,$cm$^{-1}$~\cite{RvachovPRL17}. The spin-rotation coupling between the total electronic spin and molecular rotation is expected to be smaller than 0.005$\,$cm$^{-1}$~\cite{SemczukPRL14,SemczukPRA13}. Finally, the spin-spin coupling due to the magnetic dipolar and second-order spin-orbit interactions is expected to be smaller than 0.01$\,$cm$^{-1}$~\cite{SemczukPRA13,KozlovPCCP20}. Thus all these corrections could be neglected.

\paragraph{Summary.} We have reported the accurate calculation of the interaction energy, scattering length, rovibrational structure, and permanent electric dipole moment of the ${}^{23}$Na${}^6$Li molecule in the triplet $a^3\Sigma^+$ electronic state. We have reached spectroscopic accuracy ($<0.5\,$cm$^{-1}$) using state-of-the-art \textit{ab  initio} methods of quantum chemistry without any adjustment to experimental data. We have calculated and benchmarked the importance of higher-level excitations, core-electron correlation, relativistic, QED, and adiabatic corrections, and their theoretical uncertainties to reproduce accurately scattering and spectroscopic properties. We have demonstrated that not only the leading relativistic effects but also the leading QED and adiabatic corrections are essential. We have predicted the dissociation energy and the scattering length in good agreement with experimental data. Finally, we have shown that the recent experimental potential obtained as a fit to the highly accurate vibrational spectrum~\cite{RvachovPCCP18b} had used an inaccurate theoretical equilibrium distance. Therefore, the present theoretical potential has a more accurate shape and we recommend using the present equilibrium distance or the present potential to calculate the rotational spectrum. 

The feasible improvements of theoretical methods and computational power may allow to directly test many-electron relativistic and QED theory by comparing theoretical predictions to results of scattering experiments and precision spectroscopy with ultracold molecules in the future. Potential deviations of theoretical calculations from experimental measurements would indicate the need for the development of the relativistic and QED theory beyond existing techniques for several-electron molecules.

\begin{acknowledgments} We want to thank Bogumi{\l} Jeziorski and Krzysztof Pachucki for useful comments on the manuscript. Financial support from the National Science Centre Poland (2015/19/D/ST4/02173, 2016/23/B/ST4/03231) and the Foundation for Polish Science within the First Team programme cofinanced by the European Union under the European Regional Development Fund is gratefully acknowledged. The computational part of this research has been partially supported by the PL-Grid Infrastructure.
\end{acknowledgments}

\bibliography{NaLi}

\end{document}